\documentclass[12pt,preprint]{aastex}

\shorttitle{Two types of softening}

\shortauthors{Qin et al.}

\begin{document}

\title{Two types of softening detected in X-ray
afterglows of Swift bursts: internal and external shock origins?}

\author{Y.-P. Qin\altaffilmark{1,2}, A. C. Gupta\altaffilmark{3,1},
J. H. Fan\altaffilmark{1}, R.-J. Lu\altaffilmark{2}}

\altaffiltext{1}{Center for Astrophysics, Guangzhou University,
Guangzhou 510006, P. R. China; ypqin@gzhu.edu.cn}

\altaffiltext{2}{Physics Department, Guangxi University, Nanning
530004, P. R. China}

\altaffiltext{3}{Aryabhatta Research Institute of Observational
Sciences (ARIES), Manora Peak, Nainital - 263129, India}

\begin{abstract}

The softening process observed in the steep decay phase of early
X-ray afterglows of Swift bursts has remained a puzzle since its
discovery. The softening process can also be observed in the later
phase of the bursts and its cause has also been unknown. Recently,
it was suggested that, influenced by the curvature effect, emission
from high latitudes would shift the Band function spectrum from
higher energy band to lower band, and this would give rise to the
observed softening process accompanied by a steep decay of the flux
density. The curvature effect scenario predicts that the terminating
time of the softening process would be correlated with the duration
of the process. In this paper, based on the data from the UNLV GRB
group web-site, we found an obvious correlation between the two
quantities. In addition, we found that the softening process can be
divided into two classes: the early type softening ($t_{s,max}\leq
``4000"s$) and the late type softening ($t_{s,max} > ``4000"s$). The
two types of softening show different behaviors in the duration vs.
terminating time plot. In the relation between the variation rates
of the flux density and spectral index during the softening process,
a discrepancy between the two types of softening is also observed.
According to their time scales and the discrepancy between them, we
propose that the two types are of different origins: the early type
is of internal shock origin and the late type is of external shock
origin. The early softening is referred to the steep decay just
following the prompt emission, whereas the late decay typically
conceives the transition from flat decay to late afterglow decay. We
suspect that there might be a great difference of the Lorentz factor
in two classes which is responsible for the observed discrepancy.

\end{abstract}

\keywords{gamma-rays: bursts --- gamma-rays: theory ---
relativity}

\section{Introduction}

The newly discovered phenomenon, the {\it spectral softening} of
gamma-ray bursts (GRBs) observed by the Swift instruments in their
early X-ray afterglows [1$-$5] has not been predicted by any of the
existing GRB models. Credit of this discovery goes to the Swift
instruments, which can monitor X-ray emission of GRBs at quite early
time after the trigger events. Soon after this discovery, authors of
Ref. [6] performed a systematic analysis on a selected sample of
Swift bursts and found that $\sim 75\%$ of the bursts (33 out of 44)
show an obvious spectral softening process. Besides their papers,
the UNLV (University of Nevada, Las Vegas) GRB group also presented
the temporal and spectral data of Swift bursts on their
web-site\footnote{http://swift.physics.unlv.edu/.}, which are
continued to be accumulated as the number of Swift bursts keeps
increasing. Their data show, besides the steep decay phase, the
softening can be observed in later phases as well (see our analysis
below).

Most softening processes are detected in the steep decay phase in
the early X-ray afterglows of GRBs. Since the steep decay phase
promptly follows, and is smoothly connected to, the prompt emission
phase, it is regarded as the prompt emission tail [7$-$10]. It is
generally believed that this phenomenon is due to the high latitude
emission of fireballs, in which the so-called curvature effect must
play a role [6, 9, 11$-$18]. However, the steep decay tails are
expected from the curvature effect but the softening process is not,
which puzzled astronomers.

The curvature effect arises from the emission from the surface of an
relativistically expanding fireball, where the delay of time, the
shifting of the intrinsic spectrum due to high latitude emission
areas, as well as other relevant factors of expanding fireballs must
be taken into account (for detail explanation and analysis, see [17,
19$-$21]). Due to the great amount of energy release,
relativistically expanding fireballs would be produced at very early
epoch of GRBs [22, 23]. The curvature effect is thus expected in the
prompt gamma-ray emission phase. Investigations on the profile of
the light curves of pulses, the spectral lags, the power-law
relation between the pulse width and the energy, the evolution of
the hardness ratio and the evolution of the peak energy in the
prompt emission phase have been performed by various groups in the
last decade [19$-$21, 24$-$34]. It was shown that the effect can
also play an important role in the early afterglow period [11, 12,
17, 35].

There are only few attempts of interpreting the softening phenomenon
which indicate that the phenomenon is beyond the expectation of
current or underlying models. The few attempts of interpretation
include: cooling of the internal-shocked region might be responsible
for strong softening [6]; at least in some bursts, the softening
might be accounted for by the central engine, which is assumed to
produce a soft and decaying afterglow emission [1, 6, 36]; both the
temporal behavior and the spectral softening of bursts might be a
consequence of the cannonball model of GRBs [37]; a hard-to-soft
behavior lasting to the latest phases of the afterglow can be
expected base on the ``fireshell'' model with a ``canonical GRB''
light curve containing two sharply different components [38, 39].
One of the most remarkable investigation on this issue is performed
by authors of Ref. [16]. They concluded that the early emission in
$>90\%$ of early afterglows has a curved $\nu f_{\nu}$ spectrum and
that $E_{peak}$ (peak energy) likely evolves from the $\gamma$-rays
through the soft X-ray bands on the timescales of $10^2 - 10^4$ s
after the GRBs. Along with this is the discovery of Ref. [5]: the
$E_{peak}$ of GRB 060614, which is one of the members of Ref. [6]'s
sample, decreases to as low as $\sim$ 8 keV at the beginning of the
XRT observations. The same phenomenon was revealed in literature as
early as in 2000 by the analysis of BeppoSAX data: the peak energy
was found to evolve from the prompt to the afterglow phase of GRBs,
decreasing from $>700 keV$ to $<3 keV$ for some bursts [40].

Motivated by Ref. [16]'s finding, very recently, the author of [35]
has shown that the curvature effect alone can produce both the
softening and the decaying behavior observed in the early X-ray
afterglow of the Swift bursts. It is due the shifting of the Band
function spectrum [41] which gives rise to the softening along with
the temporal decaying. Two factors of the curvature effect, the time
delay and the variation of the Doppler effect of higher latitude
emission from the fireball surface, cause the shifting of the Band
function spectrum.

According to the curvature effect scenario, the start time of the
softening is much smaller than the terminating time of the process,
and then it is expected that the softening duration must be
correlated with its terminating time [35]. We will investigate
statistically in the following if the two quantities are correlated
or not. At the same time, some other statistical properties will
also be explored. We will not limit our analysis on the steep decay
phase, but instead, any softening detected in the XRT light curve
will be considered.

The paper is organized as follows: in Section 2, we describe our
sample of Swift data and the data reduction; in Section 3, we
discuss the statistical properties obtained from our analysis;
conclusions of the present work are reported in the last section.

\section{Data}

The data (up to May 23, 2008) employed in our analysis on the
relation between the duration and terminating time of the softening
process are taken from the UNLV GRB group web-site (see footnote 1),
where Swift/XRT time-resolved spectra of selected bursts are
available (see Ref. [6] for selection criteria of the bursts). The
softening is observed in some of the bursts and we selected only
those bursts which have noticeable X-ray softening, i.e., those
bursts which should contain at least three data points starting from
a smaller spectral index and ending at a larger one (refer to the
time intervals presented in Table 1 and the corresponding spectral
evolution figures presented in the UNLV GRB group web-site).

The start time ($t_{s,min}$) and the terminating time ($t_{s,max}$),
together with the corresponding values of the spectral index
($\beta_{min}$ and $\beta_{max}$), of the softening process of the
selected bursts are listed in Table 1. Here we divide the softening
process into two distinct classes according to the corresponding
terminating time: for class 1, $t_{s,max}\leq ``4000"s$ (called the
early type softening); and for class 2, $t_{s,max} > ``4000"s$
(called the late type softening). Note that, for some bursts, there
might exist both types of softening (see Table 1).

\begin{figure}[tbp]
\begin{center}
\includegraphics[width=5in,angle=0]{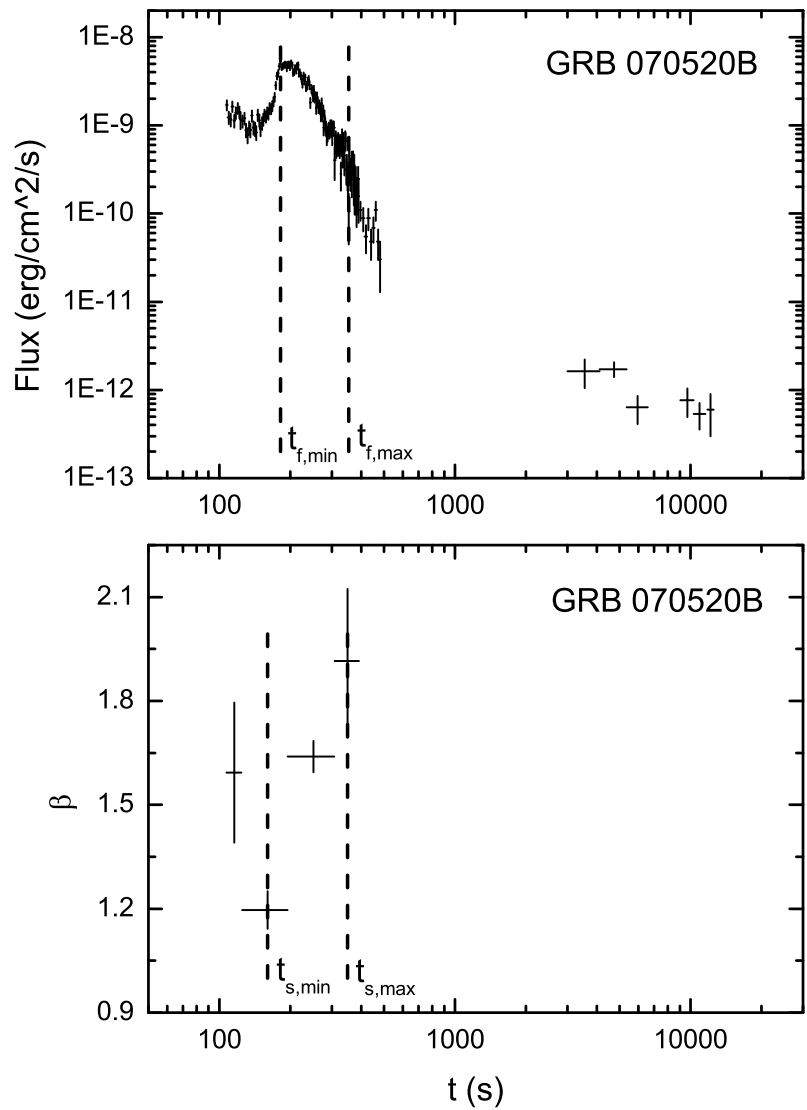}
\end{center}
\caption{The spectral evolution (the lower panel) and light curve
(the upper panel) of GRB 070520B. The data are taken from
http://swift.physics.unlv.edu/ (see the spectra.txt and the lc.txt
files of GRB 070520B). The dash lines in the lower panel denote the
time positions of $t_{s,min}$ and $t_{s,max}$, and those in the
upper panel represent that of $t_{f,min}$ and $t_{f,max}$.}
\label{Fig. 1}
\end{figure}

An examples of selecting the softening as well as the corresponding
start time and terminating time is displayed in the lower panel of
Fig. 1 (note that we employ index $\beta$ instead $\Gamma$), where
the spectral evolution of GRB 070520B is shown. The light curve of
this burst is displayed in the upper panel of the figure. The
spectral and light curve data are taken from
http://swift.physics.unlv.edu/ (spectra.txt and lc.txt files). We
first selected $t_{s,min}$ and $t_{s,max}$ by viewing the $\beta$
vs. $t$ plot (see the lower panel of Fig. 1 and the dash lines
there), and then located them in the spectra.txt file of the burst,
and then read and calculated (see the explanation below) their
values as well as their uncertainties from this file. The provided
data have been analyzed by the UNLV GRB group. The details of the
analysis are described in Ref. [6]. The UNLV GRB group have
developed a time filter for the time-resolved spectral analysis
which can be automatically performed. Time intervals for analyzing
the spectral index are determined by two criteria raised by them,
and hence they are different from each other. For example, for GRB
070520B, the lower limit of the time interval associated with its
$\beta_{min}=1.196\pm 0.054$ is $t_1 = 125 s$ and the upper limit of
this interval is $t_2 = 195 s$  (see Fig. 1 and Table 1, and also
the spectra.txt file). This gives rise to a time interval of $70 s$.
According to the data format file, the mean time of this interval is
$t=(t_1+t_2)/2=160 s$ and its error is $\sigma_t=(t_2-t_1)/2=35 s$
(see Table 1). However, the lower limit of the time interval
associated with $\beta_{max}=1.92\pm 0.21$ of this burst is $t_1 =
308 s$, and the upper limit of this interval is $t_2 = 391 s$. That
measures the time interval as $83 s$. The mean time of this interval
is $t=(t_1+t_2)/2=349.5 s$ and its error is
$\sigma_t=(t_2-t_1)/2=41.5 s$ (see Table 1). The UNLV GRB group use
different NH for different sources. They used the XSPEC spectral
fitting model: a simple power law combined with the absorptions of
both our Galaxy and the GRB host galaxy, wabs$^{Gal} \times$
zwabs$^{host} \times$ power law (for bursts with known redshifts) or
wabs$^{Gal} \times$ wabs$^{host} \times$ power law (for bursts whose
redshifts are unknown) (see Ref. [6]).

\section{Results}

\subsection{Relation between the duration and terminating time of the softening process}

\begin{figure}[tbp]
\begin{center}
\includegraphics[width=5in,angle=0]{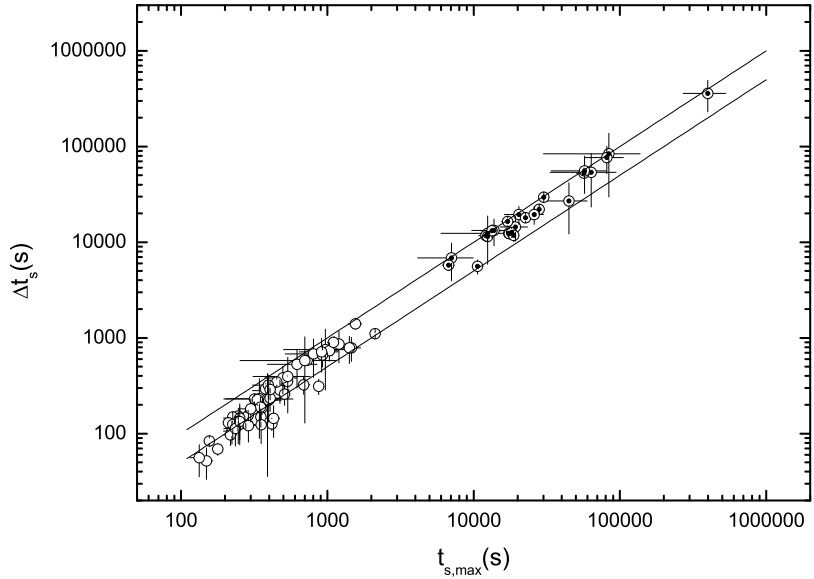}
\end{center}
\caption{Relation between the duration $\Delta t_s$ and the
 terminating time $t_{s,max}$ of the softening process observed in the
X-ray afterglows of the selected bursts. The upper and lower solid
lines are drawn by $\Delta t_s = t_{s,max}$ and $\Delta t_s =
t_{s,max}/2$, respectively. Open circles and open circles with dots
represent the early type softening and the late type softening,
respectively.} \label{Fig. 1}
\end{figure}

The duration of the softening is calculated by $\Delta t_{s} =
t_{s,max} - t_{s,min}$. The relation between $\Delta t_s$ and
$t_{s,max}$ for these bursts is displayed in Fig. 2. We find that
the data of ($t_{s,max}$, $\Delta t_s$) for the selected bursts are
distributed mainly within the area confined by the lines represented
by $\Delta t_s = t_{s,max}$ and $\Delta t_s = t_{s,max}/2$. We
observed a difference between the two types of softening: the
($t_{s,max}$, $\Delta t_s$) data for the late type softening are
well within the mentioned area, while some of the data for the early
type softening dropped out the mentioned area (especially when
$t_{s,max}$ is relatively smaller).

We performed a Spearman's correlation analysis and obtained: for the
early type softening, log $\Delta t_s = (1.172 \pm 0.051)$ log
$t_{s,max} - (0.69 \pm 0.14)$, and the fitting parameters are
$R=0.951$ (the correlation coefficient), $N=57$ (the number of data
points), and $P=8.64\times 10^{-30}$ (the chance probability); for
the late type softening, log $\Delta t_s = (1.026 \pm 0.040)$ log
$t_{s,max} - (0.19 \pm 0.18)$, with $R=0.982$, $N=26$, and
 $P=7.51\times 10^{-19}$. This is well in agreement with the prediction made by
the curvature effect [35]. In addition, the results show that the
correlation between the two quantities for the late type softening
well follows the trend of the identical curve, while in some extent
it betrays the identical curve for the early type softening (see
also Fig. 1). The cause of this difference is currently not known.

\subsection{Two types of softening distinguished in other aspects}

In addition to the duration and terminating time of the softening
process, we also measured the variations of the spectral index and
the flux density during this period.

The period of the softening is confined by the lower limit ($t_1$)
of the interval that measures $\beta_{min}$ and the upper limit
($t_2$) of the interval that measures $\beta_{max}$. For example,
for GRB 070520B, the lower limit of the time interval associated
with its $\beta_{min}=1.196\pm 0.054$ is $t_1 = 125 s$, and the
upper limit of the time interval associated with its
$\beta_{max}=1.92\pm 0.21$ is $t_2 = 391 s$ (see the last section),
and thus the softening period of this burst is the time interval
from $125 s$ to $391 s$. Within this period, we search the minimum
and maximum of the flux $f_{\nu}$ from the lc.txt file of GRB
070520B. The two extreme values of the flux are denoted by
$f_{\nu,min}$ and $f_{\nu,max}$ respectively, and the corresponding
times are denoted by $t_{f,min}$ and $t_{f,max}$ respectively. The
values of $t_{f,min}$ and $t_{f,max}$ and their uncertainties are
estimated in the same way adopted in measuring $t_{s,min}$ and
$t_{s,max}$ and their uncertainties (see the last section). For GRB
070520B, we found $f_{\nu,max}=(52.7\pm 3.9)\times 10^{-10} erg\cdot
cm^{-2}\cdot s^{-1}$, $f_{\nu,min}=(10.5\pm 6.1)\times 10^{-11}
erg\cdot cm^{-2}\cdot s^{-1}$, $t_{f,max}=181.6\pm 1.0 s$, and
$t_{f,min}=353.6\pm 1.0 s$ (see Table 2). One might observe that for
this burst $t_{f,max}$ is different from $t_{s,min}$ and $t_{f,min}$
is different from $t_{s,max}$. Two facts cause this difference. The
first is that time intervals for measuring the flux are generally
smaller than those for measuring the spectral index (see Fig. 1,
where the number of data points of the flux within the softening
period is much larger than that of the spectral index). The second
is that in the softening process some bursts undergo a rise phase
and then a decay phase in its light curve, and close to the lower
limit of the softening we measure the peak of the flux instead of
the flux at the very beginning of the softening process (see also
Fig. 1). (Note that we identified a softening process according to
the spectral index but not the flux.)

In Table 2, we listed the maximum and minimum values of the flux
density detected during the softening process. We calculated the
variation rates of the two quantities by $\Delta \beta / \Delta t_s
= (\beta_{max}-\beta_{min}) / \Delta t_s$ and $\Delta \rm{log}
f_{\nu} / \Delta t_f = (\rm{log} f_{\nu,max} - \rm{log} f_{\nu,min})
/ \Delta t_f$, where $\Delta t_f = t_{f,min}-t_{f,max}$ is the time
interval between $f_{\nu,max}$ and $f_{\nu,min}$, which can be
different from $\Delta t_s$ (see Table 2).

\begin{figure}[tbp]
\begin{center}
\includegraphics[width=5in,angle=0]{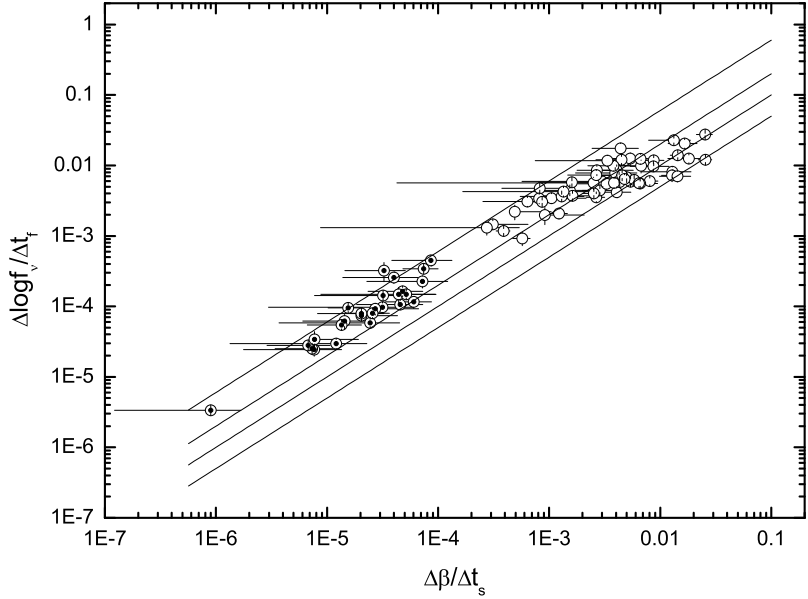}
\end{center}
\caption{Relation between the variation rates of the flux density
and spectral index, $\Delta \rm{log} f_{\nu} / \Delta t_f$ and
$\Delta \beta / \Delta t_s$, during the softening process observed
in the X-ray afterglows of the selected bursts. Solid lines from top
to the bottom represent the $\Delta \rm{log} f_{\nu} / \Delta t_f =
k \Delta \beta / \Delta t_s$ lines with $k=6$, 2, 1, and 0.5,
respectively. The open circles represent the early type softening
while the open circles with dots represent the late type softening.}
\label{Fig. 2}
\end{figure}

In Fig. 3, we presented the relation between the two variation
rates. It shows that the two types of softening do have distinct
behaviors in the variation rate. Data of the late type softening are
distributed within the $\Delta \rm{log} f_{\nu} / \Delta t_f = 2
\Delta \beta / \Delta t_s$ and $\Delta \rm{log} f_{\nu} / \Delta t_f
= 6 \Delta \beta / \Delta t_s$ curves, which well follow the trend
of the identical curve. For the early type softening, the data are
scattered in a wider area confined by the $\Delta \rm{log} f_{\nu} /
\Delta t_f = 0.5 \Delta \beta / \Delta t_s$ and $\Delta \rm{log}
f_{\nu} / \Delta t_f = 6 \Delta \beta / \Delta t_s$ curves. The
trend of the early type is obviously deviated from that of the
identical curve.

A Spearman's correlation analysis between the two variation rates
was also performed. For the early type softening we got $\rm{log}
\Delta \rm{log} f_{\nu} / \Delta t_f =  (0.556 \pm 0.045) \rm{log}
\Delta \beta / \Delta t_s - (0.85 \pm 0.11)$, with $R=0.857$,
$N=57$, and $P=1.84\times 10^{-17}$, and for the late type softening
we obtained $\rm{log} \Delta \rm{log} f_{\nu} / \Delta t_f = (0.987
\pm 0.071) \rm{log} \Delta \beta / \Delta t_s + (0.51 \pm 0.33)$,
with $R=0.943$, $N=26$, and $P=5.49\times 10^{-13}$. The correlation
must owe to the fact that the softening scope (represented by
$\Delta \beta$) and the decaying scale (described by $\Delta
\rm{log} f_{\nu}$) vary mildly for different sources, but the time
interval of the process differ significantly. The two types of
softening occupy distinct areas in the plot due to the large
variance of the softening duration between them. But why the trends
of the relation for the two types are so different remains unclear.

\section{Discussion and conclusions}

In the present work, we studied the relation between the duration
and terminating time of the softening process observed in the X-ray
afterglows of the Swift bursts. We found that these two quantities
are obviously correlated, as expected by the curvature effect. The
analysis reveals that the softening can be divided into two classes
merely on the basis of the corresponding terminating time: the early
type softening ($t_{s,max}\leq ``4000"s$) and the late type
softening ($t_{s,max} > ``4000"s$). The two types of softening show
different behaviors in their duration and the terminating time plot.
We also investigated the relation between the variation rates of the
flux density and the spectral index during the softening process. In
this aspect, more obvious discrepancy is observed between the two
types of softening.

As revealed in Ref. [35], the duration of the softening can be
affected by three parameters: the Lorentz factor, the radius, and
the intrinsic radiative peak energy concerned. As shown in Ref. [35]
Fig. 6, for the same Lorentz factor $\Gamma$ and radius $R_c$, a
smaller intrinsic peak energy $E_{0,p}$ can lead to a smaller
duration of the process, while the corresponding terminating time
will be unchanged. It would give rise to the betray observed in Fig.
1. Why this occurs in the early type softening but not in the late
type softening? We suspect, probably the early type softening is of
the internal shock origin while the late type softening is of the
external shock origin. In the former case the Lorentz factor is
large and then the curvature effect is sensitive to the fireball
parameters while in the latter case the Lorentz factor is small and
hence the curvature effect is less sensitive to the fireball
parameters. Is it due to the selection effect? This is unlikely,
because if the earlier softening tends to have a relatively smaller
duration due to the overlapping of its start time by other
components of emission then it will affect both variation rates of
the flux density and spectral index in the same way and then the
possible influence will be canceled. But Fig. 2 clearly shows that
the early softening does have a different behavior relative to the
rest.

The fact that the GRBs with lowest $t_{s,max}$ tend to deviate from
the $\Delta t_s \sim t_{s,max}$ law may be partially due to the fact
that $\Delta t_s$ is defined as the difference of the linear
quantities $t_{s,max}$ and $t_{s,min}$ and we take the logarithm of
$\Delta t_s$ and correlate it with the logarithm of $t_{s,max}$. Is
it better that we use the ratio between $t_{s,max}$ and $t_{s,min}$
rather than use $\Delta t_s$? The ratio, when passing to logarithms,
becomes the difference between $log t_{s,max}$ and $log t_{s,min}$.
Is it a more suitable quantity to be correlated with $log
t_{s,max}$? We studied this issue by replacing $\Delta t_s$ with
$t_{s,max}/t_{s,min}$ in Fig. 2. We found a weak correlation between
$log (t_{s,max}/t_{s,min})$ and $log t_{s,max}$ (the plot is
omitted). The data are quite scattered in the $log
(t_{s,max}/t_{s,min})$ vs. $log t_{s,max}$ plane. The analysis does
not provide any information of the causes of the mentioned
deviation. This must be due to the fact that $log (t_{s,max} -
t_{s,min})$ is closer to $log t_{s,max}$ than $log t_{s,max} - log
t_{s,min}$ is, as long as the discrepancy between $t_{s,max}$ and
$t_{s,min}$ is large enough (say, when $t_{s,max}$ is one order of
magnitude larger than $t_{s,min}$).

The strongest reason in favor of our suggestion of two types of
origin might be that the two types of softening occur at very
different time scales (the former appears much earlier and the
latter emerges very late). In addition we found that the early type
softening is observed in the steep decay phase which is believed to
be due to the high latitude emission of the prompt phase [6, 9,
11$-$18] and the late type softening is found in the normal
afterglow phase which was believed to be due to the external shocks
[42$-$44].

\begin{figure}[tbp]
\begin{center}
\includegraphics[width=5in,angle=0]{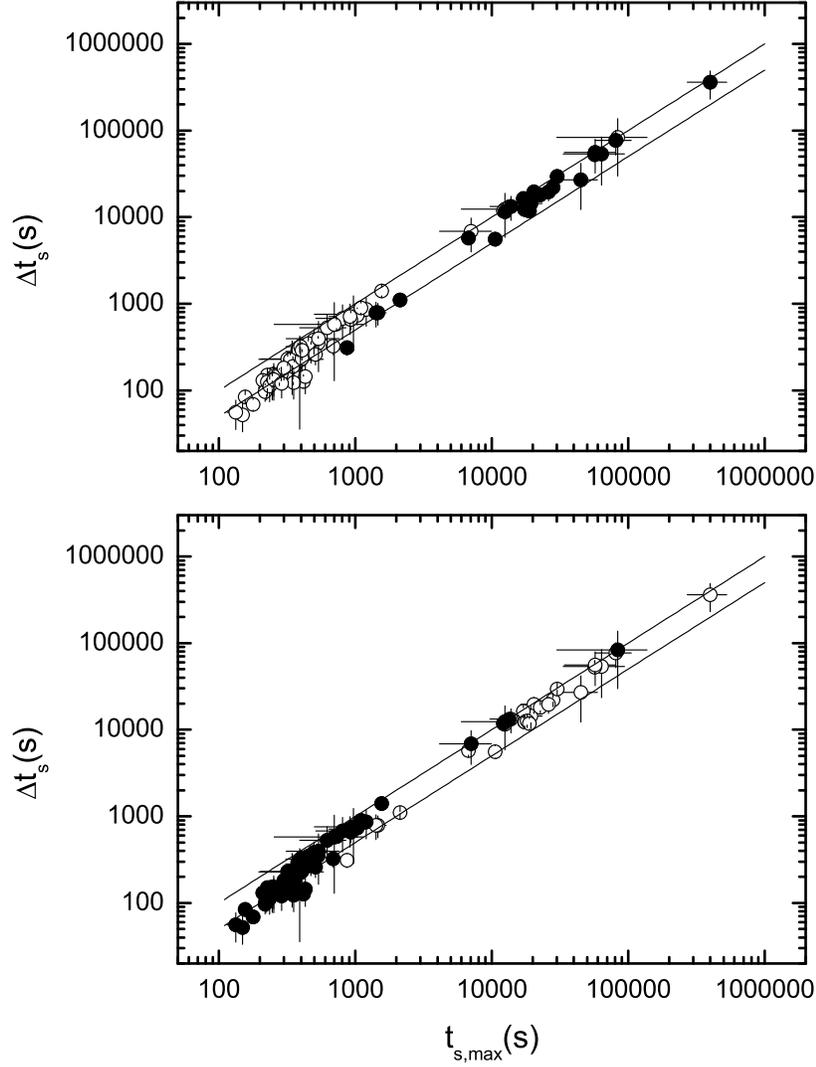}
\end{center}
\caption{Relation between $\Delta t_s$ and $t_{s,max}$ for the newly
defined two types of the softening process. Filled circles in the
upper panel and open circles in the lower panel represent the
process with $t_{s,min}>500s$ (the newly defined late type softening
process). Filled circles in the lower panel and open circles in the
upper panel stand for the process with $t_{s,min}\le 500s$ (the
newly defined early type softening process). The two solid lines are
the same as they are in Fig. 2.} \label{Fig. 1}
\end{figure}

It should be pointed out that the softening process is divided into
two classes empirically. If the two kinds of softening are
associated with internal and external shock mechanisms, it might be
more natural to divide them according to the start time of the
process. By examining the data in Tables 1 and 2, we roughly
redivided the softening process into two classes according to
$t_{s,min}$ being less or larger than $500 s$. Shown in Fig. 4 are
the relations between the duration and the terminating time of the
softening process for the newly defined early and late types. As
expected, they do not dramatically affect the result. The two types
are distributed in two distinguishable domains in the $\Delta t_s$
vs. $t_{s,max}$ plane. However, a slight overlapping between the two
distributions is observed. We do not know if this overlapping is due
to the overlapping of physical parameters or merely the statistical
fluctuation. One might notice that few of the new early type occupy
the late type domain near the identical curve and few of the late
type are located in the early type domain near the $\Delta t_s =
t_{s,max}/2$ curve. Both seem to be a result of the extension of the
two types. We therefore insist that this possibility cannot be ruled
out with the current data. However, as an empirical analysis, we
prefer the former division, i.e. dividing them according to
$t_{s,max}$, since no overlapping is observed in this division
scenario.

\begin{figure}[tbp]
\begin{center}
\includegraphics[width=5in,angle=0]{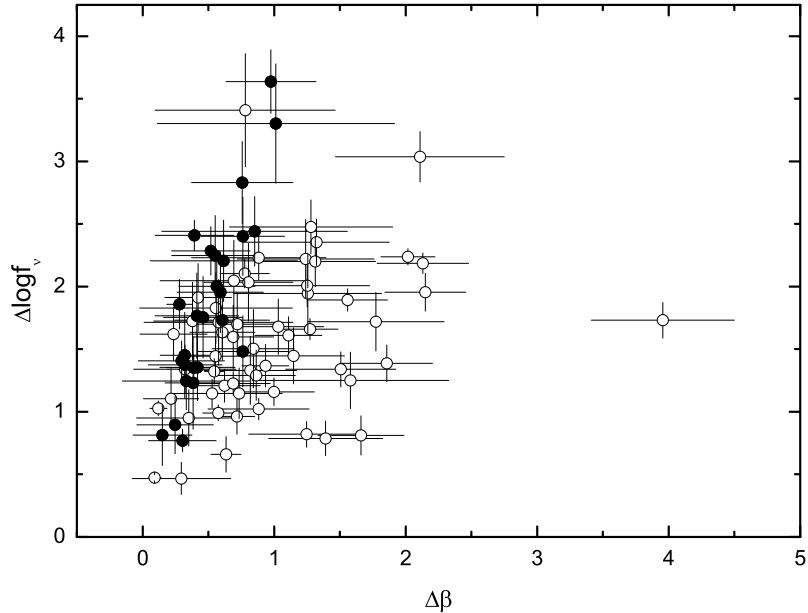}
\end{center}
\caption{Relation between the variations of the flux density and
spectral index, $\Delta \rm{log} f_{\nu}$ and $\Delta \beta$, during
the softening process observed in the X-ray afterglows of the
selected bursts. Filled circles stand for the late type of softening
and open circles for the early type.} \label{Fig. 2}
\end{figure}

Studied in Fig. 3 is the relation between the variation rates of the
flux density and spectral index during the softening process. How it
would be when we simply study the relation between $\Delta \rm{log}
f_{\nu}$ and $\Delta \beta$? Shown in Fig. 5 is the result. One
finds that the two types seem to have different distributions of
$\Delta \rm{log} f_{\nu}$ and $\Delta \beta$. The late type
softening tends to have smaller $\Delta \beta$ and slightly larger
$\Delta \rm{log} f_{\nu}$. However, the overlapping is so heavy that
we cannot tell the type of a softening merely according to its
location in the $\Delta \rm{log} f_{\nu}$ vs. $\Delta \beta$ plane.
This difference, if confirmed statistically later, might become a
hint in searching the physical difference between the two softening
processes.

\acknowledgments

Our special thanks are given to the anonymous referee for his or her
comments and suggestions which have improved the paper greatly. This
work is supported in part by the National Natural Scientific
Foundation of China (10573005, 10633010, 10747001) and the 973
project (No. 2007CB815405). We also thank the financial support from
the Guangzhou Education Bureau and Guangzhou Science and Technology
Bureau.

%\begin{thebibliography}{}
%\bibliography{}

\section{References}

[1]{} ~S. Campana et al., 2006, {\it Nature}, 442, 1008

[2]{} ~G. Ghisellini et al., 2006, {\it Mon. Not. R. Astron. Soc}, 372, 1699

[3]{} ~N. Gehrels et al., 2006, {\it Nature}, 444, 1044

[4]{} ~B. Zhang et al., 2007, {\it Astrophys. J.}, 655, L25

[5]{} ~V. Mangano et al., 2007, {\it Astron. Astrophys.}, 470, 105

[6]{} ~B. B. Zhang, E. W. Liang, and B. Zhang, 2007, {\it Astrophys. J.}, 666, 1002

[7]{} ~G. Tagliaferri et al., 2005, {\it Nature}, 436, 985

[8]{} ~S. D. Barthelmy et al., 2005, {\it Astrophys. J.}, 635, L133

[9]{} ~E. W. Liang et al., 2006, {\it Astrophys. J.}, 646, 351

[10]{} ~T. Sakamoto et al., 2007, {\it Astrophys. J.}, 669, 1115

[11]{} ~P. Kumar, and A. Panaitescu, 2000, {\it Astrophys. J.}, 541, L51

[12]{} ~C. D. Dermer, 2004, {\it Astrophys. J.}, 614, 284

[13]{} ~J. Dyks et al., 2005, astro-ph/0511699

[14]{} ~A. Panaitescu et al., 2006, {\it Mon. Not. R. Astron. Soc}, 366, 1357

[15]{} ~B. Zhang et al., 2006, {\it Astrophys. J.}, 642, 354

[16]{} ~N. R. Butler, and D. Kocevski, 2007, {\it Astrophys. J.}, 663, 407

[17]{} ~Y.-P. Qin, 2008, {\it Astrophys. J.}, 683, 900

[18]{} ~R. L. C. Starling et al., 2008, {\it Mon. Not. R. Astron. Soc}, 384, 504

[19]{} ~Y.-P. Qin, 2002, {\it Astron. Astrophys.}, 396, 705

[20]{} ~Y.-P. Qin et al., 2004, {\it Astrophys. J.}, 617, 439

[21]{} ~Y.-P. Qin et al., 2006, {\it Phys. Rev. D} 74, 063005

[22]{} ~J. Goodman, 1986, {\it Astrophys. J.}, 308, L47

[23]{} ~B. Paczynski, 1986, {\it Astrophys. J.}, 308, L43

[24]{} ~E. E. Fenimore et al., 1996, {\it Astrophys. J.}, 473, 998

[25]{} ~R. Sari, and T. Piran, 1997, {\it Astrophys. J.}, 485, 270

[26]{} ~F. Ryde, and V. Petrosian, 2002, {\it Astrophys. J.}, 578, 290

[27]{} ~D. Kocevski, F. Ryde, and E.-W. Liang, 2003, {\it Astrophys. J.}, 596, 389

[28]{} ~Y.-P. Qin, and R.-J. Lu, 2005, {\it Mon. Not. R. Astron. Soc}, 362, 1085

[29]{} ~R.-F. Shen et al., 2005, {\it Mon. Not. R. Astron. Soc}, 362, 59

[30]{} ~R.-J. Lu, Y.-P. Qin, Z.-B. Zhang, and T.-F. Yi, 2006,
{\it Mon. Not. R. Astron. Soc}, 367, 275

[31]{} ~R.-J. Lu, Z.-Y. Peng, and W. Dong, 2007, {\it Astrophys. J.}, 663, 1110

[32]{} ~Z.-Y. Peng et al., 2006, {\it Mon. Not. R. Astron. Soc}, 368, 1351

[33]{} ~Y.-P. Qin et al., 2005, {\it Astrophys. J.}, 632, 1008

[34]{} ~L.-W. Jia, 2008, {\it Chinese J. Astron. Astrophys.}, 4, 451

[35]{} ~Y.-P. Qin, 2008, ApJ, accepted [arXiv:0806.3339v1]

[36]{} ~Y. Z. Fan et al., 2006, {\it JCAP}, 9, 13

[37]{} ~S. Dado, A. Dar, and A. D. Rujula, 2008, {\it Astrophys.
J.}, 681, 1408

[38]{} ~R. Ruffini et al., 2006, {\it Astrophys. J.}, 645, L109

[39]{} ~C. L. Bianco et al., 2008, {\it AIPC}, 966, 12

[40]{} ~F. Frontera et al., 2000, {\it Astrophys. J. Suppl. Ser.},
127, 59

[41]{} ~D. Band et al. 1993, {\it Astrophys. J.}, 413, 281

[42]{} ~P. Meszaros, and M. J. Rees, 1997, {\it Astrophys. J.}, 476,
232

[43]{} ~R. Sari, T. Piran, and R. Narayan, 1998, {\it Astrophys.
J.}, 497, L17

[44]{} ~R. A. Chevalier, and Z.-Y. Li, 2000, {\it Astrophys. J.},
536, 195

%\end{thebibliography}

\begin{deluxetable}{llllll}

\tablewidth{0pt} \tablecaption{The minimum and maximum values of the
spectral index ($\beta_{min}$ and $\beta_{max}$) and the
corresponding times ($t_{s,min}$ and $t_{s,max}$), of the softening
process of the selected bursts.} \tablehead{ \colhead{type} &
\colhead{burst} & \colhead{$t_{s,min} (s)$} &
\colhead{$\beta_{min}$} & \colhead{$t_{s,max} (s)$} &
\colhead{$\beta_{max}$} } \startdata

 early & 050315    &   91.0    $\pm$   7.0    &   1.02    $\pm$
  0.27    &   620    $\pm$   230    &   1.58    $\pm$
  0.45    \\
  & 050421    &   124.0    $\pm$   9.0    & -0.23    $\pm$
  0.32    &   247    $\pm$   44    &   1.35    $\pm$
  0.36    \\
  & 050502B   &   340    $\pm$   270    &  0.972    $\pm$
  0.053 &   1200    $\pm$   150    &   2.29    $\pm$
  0.26    \\
  & 050714B   &   188    $\pm$   31    &   4.80    $\pm$
  0.20    &   474    $\pm$   73    &   6.06    $\pm$
  0.40    \\
  & 050716    &   114    $\pm$   10    & -0.13    $\pm$
  0.12    &   396    $\pm$   88    &   1.019    $\pm$
  0.084 \\
  & 050717    &   94.0    $\pm$   3.0    &  0.08 $\pm$
  0.24    &   247    $\pm$   32    &  0.90    $\pm$
  0.20    \\
  & 050724    &   87.9    $\pm$   4.3    &  0.38    $\pm$
  0.10    &   318    $\pm$   16    &   2.24    $\pm$
  0.31    \\
  & 050726    &   366    $\pm$   32    &  0.72    $\pm$
  0.26    &   689    $\pm$   58    &   1.02    $\pm$
  0.26    \\
  & 050730    &   145    $\pm$   12    &  0.245    $\pm$
  0.089 &   729    $\pm$   66    &  0.962    $\pm$
  0.062 \\
  & 050814    &   173.1    $\pm$   8.1    &  0.885    $\pm$
  0.091 &   358    $\pm$   26    &   1.88    $\pm$
  0.25    \\
  & 050904    &   182    $\pm$   13    &  0.043 $\pm$
  0.072 &   533    $\pm$   48    &  0.977    $\pm$
  0.093 \\
  & 050922B   &   673    $\pm$   17    &  0.93    $\pm$
  0.29    &   1460    $\pm$   220    &   2.21    $\pm$
  0.41    \\
  & 051117A   &   153    $\pm$   40    &  0.722    $\pm$
  0.028 &   1560    $\pm$   170    &   1.268    $\pm$
  0.027 \\
  & 051227    &   112    $\pm$   11    &  0.21    $\pm$
  0.20    &   340    $\pm$   140    &  0.91    $\pm$
  0.31    \\
  & 060115    &   126.7    $\pm$   4.5    &  0.71    $\pm$
  0.14    &   810    $\pm$   290    &   1.60    $\pm$
  0.31    \\
  & 060124    &   559    $\pm$   31    &  0.032 $\pm$
  0.023 &   871    $\pm$   49    &   1.591    $\pm$
  0.087 \\
  & 060210    &   113.0    $\pm$   9.0    &  0.472    $\pm$
  0.093 &   500    $\pm$   120    &   1.503    $\pm$
  0.049 \\
  & 060211A   &   202    $\pm$   16    &  0.689    $\pm$
  0.058 &   351    $\pm$   29    &   1.30    $\pm$
  0.20    \\
  & 060218    &   1013    $\pm$   13    &  0.424    $\pm$
  0.087 &   2122    $\pm$   27    &   1.057    $\pm$
  0.072 \\
  & 060413    &   132    $\pm$   11    &  0.626    $\pm$
  0.099 &   278    $\pm$   31    &   1.25    $\pm$
  0.21    \\
  & 060510B   &   291    $\pm$   11    &  0.214    $\pm$
  0.060 &   417    $\pm$   16    &   1.32    $\pm$
  0.18    \\
  & 060522    &   160    $\pm$   11    &  0.40    $\pm$
  0.24    &   390    $\pm$   190    &  0.98    $\pm$
  0.16    \\
  & 060526    &   285    $\pm$   30    &  0.640    $\pm$
  0.037 &   429    $\pm$   44    &   1.89    $\pm$
  0.13    \\
  & 060607A   &   97    $\pm$   10    &  0.428    $\pm$
  0.073 &   149    $\pm$   16    &   1.12    $\pm$
  0.11    \\
  & 060607A   &   229    $\pm$   24    &  0.560    $\pm$
  0.061 &   353    $\pm$   38    &   1.114    $\pm$
  0.088 \\
  & 060614    &   104.3    $\pm$   5.5    &  0.070 $\pm$
  0.042 &   451    $\pm$   25    &   2.09    $\pm$
  0.13    \\
  & 060707    &   213    $\pm$   49    &  0.72    $\pm$
  0.15    &   970    $\pm$   470    &  0.96    $\pm$
  0.14    \\
  & 060714    &   116.5    $\pm$   7.5    &  0.361    $\pm$
  0.083 &   223    $\pm$   28    &   2.13    $\pm$
  0.17    \\
  & 060729    &   132.09    $\pm$  0.50    &  0.94    $\pm$
  0.20    &   286.8    $\pm$   7.5    &   4.90    $\pm$
  0.47    \\
  & 060814    &   83.3    $\pm$   4.8    &  0.241    $\pm$
  0.074 &   377    $\pm$   36    &   1.51    $\pm$
  0.12    \\
  & 060904A   &   76.2    $\pm$   3.0    &  0.036 $\pm$
  0.079 &   226    $\pm$   14    &   2.19    $\pm$
  0.21    \\
  & 061007    &   106    $\pm$   18    &  0.847    $\pm$
  0.032 &   252    $\pm$   57    &  0.967    $\pm$
  0.029 \\
  & 061110A   &   80.4    $\pm$   3.0    &   1.57    $\pm$
  0.14    &   210    $\pm$   15    &   3.23    $\pm$
  0.21    \\
  & 061121    &   71.7    $\pm$   2.0    & -0.455    $\pm$
  0.062 &   155.8    $\pm$   5.3    &   1.68    $\pm$
  0.31    \\
  & 061222A   &   108.8    $\pm$   2.3    &  0.80    $\pm$
  0.20    &   177.8    $\pm$   9.6    &   2.05    $\pm$
  0.35    \\
  & 070110    &   111    $\pm$   10    &  0.79    $\pm$
  0.11    &   245    $\pm$   25    &   1.32    $\pm$
  0.26    \\
  & 070129    &   299    $\pm$   46    &  0.250    $\pm$
  0.030 &   1030    $\pm$   150    &   2.36    $\pm$
  0.44    \\
  & 070223    &   121.0    $\pm$   3.0    & -0.22    $\pm$
  0.26    &   218    $\pm$   22    &   1.17    $\pm$
  0.13    \\
  & 070318    &   174    $\pm$   22    &  0.19    $\pm$
  0.12    &   397    $\pm$   89    &  0.924    $\pm$
  0.069 \\
  & 070330    &   141    $\pm$   54    &  0.43    $\pm$
  0.27    &   540    $\pm$   230    &  0.76    $\pm$
  0.20    \\
  & 070419B   &   106    $\pm$   17    &  0.569    $\pm$
  0.047 &   333    $\pm$   25    &   1.144    $\pm$
  0.076 \\
  & 070518    &   101    $\pm$   22    &   1.147    $\pm$
  0.093 &   226    $\pm$   31    &   1.99    $\pm$
  0.15    \\
  & 070520B   &   160    $\pm$   35    &   1.196    $\pm$
  0.054 &   350    $\pm$   42    &   1.92    $\pm$
  0.21    \\
  & 070616    &   204    $\pm$   65    & -0.060 $\pm$
  0.028 &   1100    $\pm$   100    &  0.72    $\pm$
  0.15    \\
  & 070621    &   122.0    $\pm$   2.0    &   1.18    $\pm$
  0.21    &   235    $\pm$   39    &   1.56    $\pm$
  0.16    \\
  & 070704    &   252    $\pm$   57    &  0.385    $\pm$
  0.084 &   511    $\pm$   26    &   1.07    $\pm$
  0.25    \\
  & 070714B   &   74.0    $\pm$   4.0    & -0.11    $\pm$
  0.22    &   396    $\pm$   88    &   1.13    $\pm$
  0.33    \\
  & 070721B   &   265    $\pm$   44    &  0.08 $\pm$
  0.16    &   930    $\pm$   160    &  0.50    $\pm$
  0.17    \\
  & 071031    &   118.0    $\pm$   6.0    &  0.320    $\pm$
  0.078 &   252    $\pm$   57    &   1.201    $\pm$
  0.035 \\
  & 071112C   &   624    $\pm$   140    &  0.43    $\pm$
  0.11    &   1410    $\pm$   210    &  0.65    $\pm$
  0.16    \\
  & 080123    &   124    $\pm$   13    &  0.57    $\pm$
  0.10    &   700    $\pm$   450    &   1.35    $\pm$
  0.30    \\
  & 080319B   &   77    $\pm$   10    &  0.671    $\pm$
  0.021 &   133    $\pm$   18    &  0.761    $\pm$
  0.022 \\
  & 080325    &   173    $\pm$   12    &  0.682    $\pm$
  0.070 &   407    $\pm$   62    &   2.188    $\pm$
  0.068 \\
  & 080430    &   196    $\pm$   54    &  0.56    $\pm$
  0.27    &   910    $\pm$   270    &  0.91    $\pm$
  0.26    \\
  & 080503    &   118    $\pm$   18    &  0.240    $\pm$
  0.057 &   407    $\pm$   62    &   1.55    $\pm$
  0.35    \\
  & 080506    &   167    $\pm$   18    &  0.224    $\pm$
  0.051 &   288    $\pm$   35    &   1.03    $\pm$
  0.21    \\
  & 080523    &   118    $\pm$   18    &   1.287    $\pm$
  0.063 &   299    $\pm$   45    &   2.15    $\pm$
  0.17    \\\hline
   late & 050315    &   40000    $\pm$   18000    &  0.99    $\pm$
  0.10    &   400000    $\pm$   130000    &   1.31    $\pm$
  0.23    \\
  & 050721    &   830    $\pm$   97    &  0.55    $\pm$
  0.22    &   20400    $\pm$   4100    &   1.16    $\pm$
  0.50    \\
  & 050726    &   4850    $\pm$   900    &  0.95    $\pm$
  0.11    &   19300    $\pm$   4100    &   1.25    $\pm$
  0.30    \\
  & 050730    &   4580    $\pm$   570    &  0.434    $\pm$
  0.049 &   22600    $\pm$   1300    &  0.714    $\pm$
  0.080 \\
  & 050803    &   9800    $\pm$   3300    &  0.879    $\pm$
  0.088 &   60000    $\pm$   30000    &   1.29    $\pm$
  0.20    \\
  & 050826    &   210    $\pm$   95    &  0.63    $\pm$
  0.30    &   84000    $\pm$   54000    &   1.64    $\pm$
  0.54    \\
  & 050904    &   17800    $\pm$   1300    &  0.730    $\pm$
  0.063 &   45000    $\pm$   15000    &   1.28    $\pm$
  0.12    \\
  & 051109A   &   4070    $\pm$   570    &  0.720    $\pm$
  0.093 &   56800    $\pm$   6200    &   1.11    $\pm$
  0.18    \\
  & 060105    &   4820    $\pm$   180    &  0.92    $\pm$
  0.15    &   81000    $\pm$   24000    &   1.44    $\pm$
  0.19    \\
  & 060204B   &   530    $\pm$   140    &   1.03    $\pm$
  0.26    &   13800    $\pm$   4100    &   1.63    $\pm$
  0.16    \\
  & 060306    &   135    $\pm$   38    &  1.00    $\pm$
  0.16    &   12500    $\pm$   6500    &   1.39    $\pm$
  0.14    \\
  & 060313    &   178    $\pm$   65    &  0.42    $\pm$
  0.16    &   7100    $\pm$   2900    &   1.01    $\pm$
  0.14    \\
  & 060714    &   980    $\pm$   220    &  0.78    $\pm$
  0.19    &   6750    $\pm$   650    &   1.19    $\pm$
  0.21    \\
  & 060807    &   4960    $\pm$   480    &   1.09    $\pm$
  0.15    &   17400    $\pm$   1300    &   1.39    $\pm$
  0.20    \\
  & 070220    &   397    $\pm$   87    &  0.23    $\pm$
  0.20    &   12100    $\pm$   1300    &  0.80    $\pm$
  0.20    \\
  & 070318    &   990    $\pm$   220    &  0.55    $\pm$
  0.11    &   12480    $\pm$   890    &   1.01    $\pm$
  0.27    \\
  & 070419B   &   5970    $\pm$   180    &  0.50    $\pm$
  0.15    &   28100    $\pm$   1200    &  0.82    $\pm$
  0.11    \\
  & 070508    &   1600    $\pm$   270    &  0.55    $\pm$
  0.12    &   57000    $\pm$   23000    &   1.31    $\pm$
  0.19    \\
  & 070721B   &   5020    $\pm$   770    &  0.43    $\pm$
  0.11    &   10610    $\pm$   530    &  0.68    $\pm$
  0.27    \\
  & 071020    &   630    $\pm$   140    &  0.53    $\pm$
  0.11    &   17000    $\pm$   1500    &   1.38    $\pm$
  0.69    \\
  & 071025    &   174    $\pm$   22    &  0.412    $\pm$
  0.049 &   13400    $\pm$   2100    &   1.39    $\pm$
  0.30    \\
  & 080207    &   5260    $\pm$   530    &   1.69    $\pm$
  0.23    &   17500    $\pm$   1400    &   2.03    $\pm$
  0.42    \\
  & 080319C   &   650    $\pm$   190    &  0.535    $\pm$
  0.079 &   30190    $\pm$   970    &   1.30    $\pm$
  0.31    \\
  & 080328    &   5590    $\pm$   200    &  0.57    $\pm$
  0.26    &   18200    $\pm$   1200    &   1.33    $\pm$
  0.20    \\
  & 080413B   &   7000    $\pm$   1000    &  0.79    $\pm$
  0.10    &   18800    $\pm$   1300    &   1.18    $\pm$
  0.18    \\
  & 080430    &   6490    $\pm$   910    &  0.75    $\pm$
  0.16    &   26000    $\pm$   4000    &  0.90    $\pm$
  0.15    \\

\enddata
\end{deluxetable}

\begin{deluxetable}{llllll}

\tablewidth{0pt} \tablecaption{The maximum and minimum values of the
flux density ($f_{\nu, max}$ and $f_{\nu, min}$) and the
corresponding times ($t_{f,max}$ and $t_{f,min}$), detected during
the softening process.} \tablehead{ \colhead{type} & \colhead{burst}
& \colhead{$t_{f,max} (s)$} & \colhead{$f_{\nu, max}^a$} &
\colhead{$t_{f,min} (s)$} & \colhead{$f_{\nu, min}^a$} } \startdata

 early
  & 050315    &   88.0    $\pm$   1.5    &  (8.6 $\pm$
  1.4)E-10 &   620    $\pm$   77    &  (12.8 $\pm$
  2.3)E-12\\
  & 050421    &   130.4    $\pm$   1.0    &  (17.1 $\pm$
  3.4)E-10 &   285.0    $\pm$   4.7    &  (9.6 $\pm$
  4.5)E-11\\
  & 050502B   &   743.2    $\pm$   1.0    &  (66.6 $\pm$
  4.6)E-10&   1277    $\pm$   26    &  (29.5 $\pm$
  9.8)E-12\\
  & 050714B   &   162.4    $\pm$   1.0    &  (19.2 $\pm$
  2.6)E-08 &   273.2    $\pm$   3.5    &  (2.2 $\pm$
  1.5)E-09\\
  & 050716    &   109.0    $\pm$   1.0    &  (25.1 $\pm$
  3.1)E-10&   457.0    $\pm$   1.0    &  (9.0 $\pm$
  4.5)E-11\\
  & 050717    &   96.0    $\pm$   1.0    &  (32.2 $\pm$
  3.9)E-10&   202.0    $\pm$   1.0    &  (15.1 $\pm$
  9.2)E-11\\
  & 050724    &   79.94    $\pm$  0.25    &  (18.6 $\pm$
  2.1)E-09 &   312.0    $\pm$   1.1    &  (7.6 $\pm$
  2.4)E-10\\
  & 050726    &   382.3    $\pm$   6.9    &  (17.3 $\pm$
  2.9)E-11 &   617.8    $\pm$   8.6    &  (5.9 $\pm$
  1.4)E-11\\
  & 050730    &   139.6    $\pm$   1.0    &  (20.5 $\pm$
  2.4)E-10 &   603.6    $\pm$   1.0    &  (22.4 $\pm$
  6.7)E-11\\
  & 050814    &   167.09    $\pm$  0.30    &  (24.6 $\pm$
  4.1)E-10 &   363.7    $\pm$   2.9    &  (17.0 $\pm$
  3.2)E-11\\
  & 050904    &   178.0    $\pm$   1.0    &  (34.6 $\pm$
  3.4)E-10 &   566.0    $\pm$   1.0    &  (14.8 $\pm$
  5.6)E-11\\
  & 050922B   &   755.6    $\pm$   1.6    &  (5.7 $\pm$
  1.0)E-09 &   1426    $\pm$   12    &  (19.1 $\pm$
  8.6)E-12\\
  & 051117A   &   134.0    $\pm$   1.0    &  (87.3 $\pm$
  5.7)E-10&   1257.7    $\pm$   1.1    &  (41.5 $\pm$
  9.3)E-11\\
  & 051227    &   114.2    $\pm$   1.0    &  (11.7 $\pm$
  2.1)E-10 &   460    $\pm$   10    &  (10.5 $\pm$
  7.4)E-12\\
  & 060115    &   124.45    $\pm$  0.25    &  (32.9 $\pm$
  6.2)E-10 &   741    $\pm$   32    &  (19.4 $\pm$
  4.6)E-12\\
  & 060124    &   572.14    $\pm$  0.25    &  (84.1 $\pm$
  5.8)E-09 &   849.50    $\pm$  0.70    &  (10.8 $\pm$
  2.1)E-10\\
  & 060210    &   106.8    $\pm$   1.0    &  (64.9 $\pm$
  5.6)E-10 &   302.8    $\pm$   1.0    &  (13.6 $\pm$
  6.8)E-11\\
  & 060211A   &   187.0    $\pm$   1.0    &  (47.1 $\pm$
  3.6)E-10 &   379.0    $\pm$   1.0    &  (10.9 $\pm$
  5.5)E-11\\
  & 060218    &   1258.04    $\pm$  0.25    &  (18.0 $\pm$
  4.0)E-09 &   1975.54    $\pm$  0.25    &  (39.3 $\pm$
  9.3)E-10\\
  & 060413    &   122.0    $\pm$   1.0    &  (12.5 $\pm$
  1.0)E-09 &   302.8    $\pm$   1.8    &  (7.7 $\pm$
  2.2)E-10\\
  & 060510B   &   305.53    $\pm$  0.25    &  (12.1 $\pm$
  1.5)E-09 &   442.59    $\pm$  0.77    &  (29.8 $\pm$
  9.4)E-11\\
  & 060522    &   158.2    $\pm$   1.0    &  (5.3 $\pm$
  1.2)E-10 &   461.4    $\pm$   7.2    &  (10.3 $\pm$
  7.7)E-12\\
  & 060526    &   255.78    $\pm$  0.61    &  (14.8 $\pm$
  1.2)E-09 &   463.4    $\pm$   1.0    &  (14.6 $\pm$
  5.5)E-11\\
  & 060607A   &   99.8    $\pm$   1.0    &  (84.9 $\pm$
  6.4)E-10 &   153.8    $\pm$   1.0    &  (50.5 $\pm$
  9.7)E-11\\
  & 060607A   &   267.8    $\pm$   1.0    &  (38.9 $\pm$
  2.8)E-10 &   387.8    $\pm$   1.0    &  (14.0 $\pm$
  4.9)E-11\\
  & 060614    &   106.03    $\pm$  0.25    &  (75.2 $\pm$
  5.6)E-09 &   465.2    $\pm$   3.6    &  (43.6 $\pm$
  5.2)E-11\\
  & 060707    &   188.3    $\pm$   3.5    &  (17.0 $\pm$
  3.3)E-11 &   1301    $\pm$   28    &  (4.1 $\pm$
  1.8)E-12\\
  & 060714    &   139.6    $\pm$   1.0    &  (72.7 $\pm$
  4.5)E-10 &   223.6    $\pm$   1.0    &  (13.9 $\pm$
  7.3)E-11\\
  & 060729    &   132.34    $\pm$  0.25    &  (11.8 $\pm$
  1.1)E-08 &   276.48    $\pm$  0.51    &  (21.9 $\pm$
  6.9)E-10\\
  & 060814    &   79.84    $\pm$  0.25    &  (45.3 $\pm$
  4.6)E-09 &   365.8    $\pm$   1.5    &  (9.9 $\pm$
  1.6)E-10\\
  & 060904A   &   74.44    $\pm$  0.25    &  (37.1 $\pm$
  3.8)E-09 &   214.86    $\pm$  0.68    &  (4.1 $\pm$
  1.4)E-10\\
  & 061007    &   91.4    $\pm$   1.0    &  (39.5 $\pm$
  1.7)E-09 &   306.96    $\pm$  0.58    &  (37.2 $\pm$
  4.8)E-10\\
  & 061110A   &   86.64    $\pm$  0.25    &  (9.9 $\pm$
  1.3)E-09&   198.37    $\pm$  0.30    &  (15.3 $\pm$
  5.1)E-10\\
  & 061121    &   74.91    $\pm$  0.25    &  (19.2 $\pm$
  1.1)E-08 &   154.29    $\pm$  0.75    &  (12.5 $\pm$
  2.2)E-10\\
  & 061222A   &   109.82    $\pm$  0.25    &  (13.0 $\pm$
  1.9)E-09 &   175.7    $\pm$   1.1    &  (19.7 $\pm$
  3.6)E-10\\
  & 070110    &   102.0    $\pm$   1.0    &  (13.8 $\pm$
  1.7)E-10 &   249.1    $\pm$   4.2    &  (9.9 $\pm$
  2.1)E-11\\
  & 070129    &   365.0    $\pm$   1.0    &  (26.5 $\pm$
  1.5)E-09 &   1054    $\pm$   10    &  (2.4 $\pm$
  1.1)E-11\\
  & 070223    &   118.9    $\pm$   1.0    &  (40.3 $\pm$
  4.9)E-10 &   231.0    $\pm$   1.0    &  (6.6 $\pm$
  1.9)E-10\\
  & 070318    &   274.8    $\pm$   1.0    &  (17.3 $\pm$
  2.2)E-10 &   482.8    $\pm$   1.0    &  (12.3 $\pm$
  5.5)E-11\\
  & 070330    &   221.9    $\pm$   2.9    &  (29.5 $\pm$
  6.1)E-11 &   616    $\pm$   16    &  (13.5 $\pm$
  5.1)E-12\\
  & 070419B   &   104.0    $\pm$   1.0    &  (15.8 $\pm$
  1.0)E-09 &   352.0    $\pm$   1.0    &  (16.1 $\pm$
  2.1)E-10\\
  & 070518    &   104.0    $\pm$   1.0    &  (12.9 $\pm$
  1.6)E-10 &   256.0    $\pm$  1.0    &  (4.0 $\pm$
  2.9)E-11\\
  & 070520B   &   181.6    $\pm$   1.0    &  (52.7 $\pm$
  3.9)E-10 &   353.6    $\pm$   1.0    &  (10.5 $\pm$
  6.1)E-11\\
  & 070616    &   486.0    $\pm$   1.0    &  (20.0 $\pm$
  1.2)E-09 &   1174.8    $\pm$   5.4    &  (15.7 $\pm$
  4.0)E-11\\
  & 070621    &   124.8    $\pm$   1.0    &  (61.9 $\pm$
  5.1)E-10 &   272.8    $\pm$   1.0    &  (11.7 $\pm$
  8.3)E-11\\
  & 070704    &   315.2    $\pm$   1.0    &  (106.5 $\pm$
  7.9)E-10 &   535.2    $\pm$   1.0    &  (2.7 $\pm$
  1.6)E-10\\
  & 070714B   &   73.5    $\pm$   1.0    &  (24.3 $\pm$
  3.2)E-10 &   466.1    $\pm$   6.0    &  (1.5 $\pm$
  1.0)E-11\\
  & 070721B   &   312.0    $\pm$   1.0    &  (25.8 $\pm$
  2.7)E-10 &   935.6    $\pm$   5.4    &  (3.1 $\pm$
  1.9)E-11\\
  & 071031    &   121.4    $\pm$   1.0    &  (58.2 $\pm$
  4.1)E-10 &   305.4    $\pm$   1.0    &  (5.5 $\pm$
  1.0)E-10\\
  & 071112C   &   569.1    $\pm$   2.4    &  (27.9 $\pm$
  6.8)E-11 &   1413.8    $\pm$   7.2    &  (2.2 $\pm$
  1.1)E-11\\
  & 080123    &   122.0    $\pm$   1.0    &  (17.7 $\pm$
  2.1)E-10 &   921    $\pm$   77    &  (6.9 $\pm$
  4.9)E-13\\
  & 080319B   &   68.2    $\pm$   1.0    &  (135.9 $\pm$
  3.4)E-09 &   150.09    $\pm$  0.91    &  (45.7 $\pm$
  2.0)E-09\\
  & 080325    &   222.2    $\pm$   1.0    &  (108.3 $\pm$
  7.4)E-10 &   460.2    $\pm$   1.0    &  (5.0 $\pm$
  1.6)E-10\\
  & 080430    &   172.3    $\pm$   6.0    &  (21.1 $\pm$
  5.3)E-11 &   603    $\pm$   15    &  (2.4 $\pm$
  1.1)E-11\\
  & 080503    &   105.2    $\pm$   1.0    &  (48.4 $\pm$
  3.9)E-10 &   409.3    $\pm$   5.0    &  (3.1 $\pm$
  1.4)E-11\\
  & 080506    &   156.4    $\pm$   1.0    &  (57.4 $\pm$
  4.4)E-10 &   320.4    $\pm$   1.0    &  (5.3 $\pm$
  3.8)E-11\\
  & 080523    &   104.8    $\pm$   1.0    &  (19.9 $\pm$
  2.0)E-10 &   306.8    $\pm$   1.0    &  (10.2 $\pm$
  4.6)E-11\\
\hline
  late
  & 050315    &   23810    $\pm$   270    &  (9.7 $\pm$
  1.1)E-12 &   432400    $\pm$   9400    &  (40.8 $\pm$
  7.4)E-14\\
  & 050721    &   777.0    $\pm$   5.0    &  (14.1 $\pm$
  3.4)E-11 &   23530    $\pm$   190    &  (8.8 $\pm$
  6.2)E-13\\
  & 050726    &   4001    $\pm$   53    &  (27.3 $\pm$
  6.6)E-12 &   23010    $\pm$   380    &  (10.7 $\pm$
  2.9)E-13\\
  & 050730    &   4370.1    $\pm$   3.7    &  (5.2 $\pm$
  1.0)E-10 &   23912    $\pm$   28    &  (7.2 $\pm$
  2.9)E-12\\
  & 050803    &   10502    $\pm$   53    &  (5.7 $\pm$
  4.0)E-11 &   83100    $\pm$   5500    &  (9.7 $\pm$
  1.4)E-13\\
  & 050826    &   169.7    $\pm$   5.0    &  (20.8 $\pm$
  5.1)E-11 &   111000    $\pm$   12000    &  (10.4 $\pm$
  6.8)E-14\\
  & 050904    &   18582.8    $\pm$   8.9    &  (8.6 $\pm$
  1.8)E-11 &   47020    $\pm$   120    &  (4.9 $\pm$
  3.4)E-13\\
  & 051109A   &   4185    $\pm$   32    &  (61.0 $\pm$
  8.3)E-12 &   57940    $\pm$   670    &  (27.2 $\pm$
  5.3)E-13\\
  & 060105    &   5571.2    $\pm$   3.4    &  (24.4 $\pm$
  5.7)E-11 &   87600    $\pm$   300    &  (12.6 $\pm$
  4.8)E-13\\
  & 060204B   &   411    $\pm$   17    &  (11.6 $\pm$
  1.9)E-11 &   16720    $\pm$   110    &  (2.1 $\pm$
  1.1)E-12\\
  & 060306    &   102.4    $\pm$   1.7    &  (10.3 $\pm$
  1.5)E-10 &   16970    $\pm$   220    &  (40.3 $\pm$
  9.5)E-13\\
  & 060313    &   149.4    $\pm$   2.4    &  (27.2 $\pm$
  6.1)E-11 &   4500    $\pm$   17    &  (3.0 $\pm$
  2.1)E-12\\
  & 060714    &   880    $\pm$   10    &  (8.4 $\pm$
  1.8)E-11 &   6860    $\pm$   28    &  (3.7 $\pm$
  2.1)E-12\\
  & 060807    &   4743    $\pm$   53    &  (36.6 $\pm$
  4.9)E-12 &   17850    $\pm$   250    &  (62.4 $\pm$
  9.5)E-13\\
  & 070220    &   542.5    $\pm$   3.5    &  (41.0 $\pm$
  8.0)E-11 &   12786    $\pm$   53    &  (4.1 $\pm$
  2.5)E-12\\
  & 070318    &   789.1    $\pm$   4.1    &  (15.8 $\pm$
  3.8)E-11 &   7584    $\pm$   39    &  (2.8 $\pm$
  2.0)E-12\\
  & 070419B   &   5807.5    $\pm$   3.8    &  (17.3 $\pm$
  4.3)E-11 &   29245    $\pm$   19    &  (6.1 $\pm$
  4.6)E-12\\
  & 070508    &   1583.3    $\pm$   2.4    &  (8.0 $\pm$
  2.1)E-10 &   53632    $\pm$   269    &  (11.7 $\pm$
  8.3)E-13\\
  & 070721B   &   4438    $\pm$   15    &  (40.4 $\pm$
  9.5)E-12 &   10501    $\pm$   38    &  (5.1 $\pm$
  2.4)E-12\\
  & 071020    &   566.6    $\pm$   2.9    &  (31.4 $\pm$
  6.8)E-11 &   17000    $\pm$   1500    &  (11.4 $\pm$
  3.6)E-13\\
  & 071025    &   159.0    $\pm$   1.0    &  (100.8 $\pm$
  7.5)E-10 &   10866    $\pm$   39    &  (2.3 $\pm$
  1.3)E-12\\
  & 080207    &   4751    $\pm$   19    &  (17.3 $\pm$
  4.1)E-11 &   18230    $\pm$   100    &  (9.8 $\pm$
  4.6)E-12\\
  & 080319C   &   505.1    $\pm$   7.7    &  (44.9 $\pm$
  6.0)E-11 &   30511    $\pm$   35    &  (1.8 $\pm$
  1.3)E-12\\
  & 080328    &   5517    $\pm$   14    &  (15.0 $\pm$
  3.4)E-11 &   18298    $\pm$   43    &  (4.9 $\pm$
  3.0)E-12\\
  & 080413B   &   8541    $\pm$   53    &  (3.5 $\pm$
  1.4)E-11 &   12367    $\pm$   53    &  (2.1 $\pm$
  1.5)E-12\\
  & 080430    &   5965    $\pm$   35    &  (14.2 $\pm$
  3.5)E-12 &   29943    $\pm$   57    &  (2.2 $\pm$
  1.1)E-12\\

\enddata
\tablenotetext{a}{in units of $erg \cdot cm^{-2} s^{-1}$.}
\end{deluxetable}

\end{document}